\documentclass[%
 reprint,
 amsmath,amssymb,
 aps,
]{revtex4-2}

\usepackage{graphicx}% Include figure files
\usepackage{dcolumn}% Align table columns on decimal point
\usepackage{bm}% bold math
\begin{document}

\preprint{APS/123-QED}

\title{Mode Control and Dynamic Population Gratings in Quantum-Dot Lasers}

\author{Xiangpeng Ou}
\affiliation{Integrated Photonics Lab (IPL), CEMSE, King Abdullah University of Science and Technology (KAUST), Thuwal 23955, Makkah Region, Saudi Arabia}

\author{Artem Prokoshin}
\affiliation{Integrated Photonics Lab (IPL), CEMSE, King Abdullah University of Science and Technology (KAUST), Thuwal 23955, Makkah Region, Saudi Arabia}

\author{Hongyan Yu}
\affiliation{Integrated Photonics Lab (IPL), CEMSE, King Abdullah University of Science and Technology (KAUST), Thuwal 23955, Makkah Region, Saudi Arabia}

\author{Xin Yao}
\affiliation{Integrated Photonics Lab (IPL), CEMSE, King Abdullah University of Science and Technology (KAUST), Thuwal 23955, Makkah Region, Saudi Arabia}

\author{Ying Shi}
\affiliation{Integrated Photonics Lab (IPL), CEMSE, King Abdullah University of Science and Technology (KAUST), Thuwal 23955, Makkah Region, Saudi Arabia}

\author{William He}
\affiliation{Integrated Photonics Lab (IPL), CEMSE, King Abdullah University of Science and Technology (KAUST), Thuwal 23955, Makkah Region, Saudi Arabia}

\author{Zhican Zhou}
\affiliation{Integrated Photonics Lab (IPL), CEMSE, King Abdullah University of Science and Technology (KAUST), Thuwal 23955, Makkah Region, Saudi Arabia}

\author{Yating Wan}
\email{yating.wan@kaust.edu.sa}
\affiliation{Integrated Photonics Lab (IPL), CEMSE, King Abdullah University of Science and Technology (KAUST), Thuwal 23955, Makkah Region, Saudi Arabia}

\date{\today}% It is always \today, today,

\begin{abstract}
Single-mode operation is essential for integrated semiconductor lasers, yet most solutions rely on regrowth, etched gratings, or other complex fabrication steps that limit scalability. We show that quantum-dot (QD) lasers can achieve stable single-mode lasing through a simple cavity design using dynamic population gratings (DPGs). Owing to the low lateral carrier diffusion of QDs, a strong standing-wave-induced carrier grating forms in a reverse-biased saturable absorber and provides self-aligned, mode-selective feedback not attainable in quantum-well devices. A single-ring  laser achieves 46 dB side-mode suppression ratio (SMSR), while a dual-ring Vernier laser delivers ($\textgreater$ 46~nm) tuning range and  up to 52.6~dB SMSR, with continuous-wave operation up to $80\,^{\circ}\mathrm{C}$. The laser remains single-mode under $-10.6$ dB external optical feedback and supports isolator-free data transmission at 32 Gbps. These results establish DPG-enabled QD lasers as a simple and scalable route to tunable, feedback-resilient on-chip light sources for communication, sensing, and reconfigurable photonic systems.
\end{abstract}

\maketitle

%\tableofcontents

\textit{Introduction}— Stable single-longitudinal-mode (SLM) operation is a core requirement in semiconductor lasers, which determines spectral purity, tuning behavior, and the stability needed for tunable sources and wavelength-selective photonic systems \cite{Mode_control, letsou2025hybridized,9609553,9187937,zhou2023prospects,Liang:16}. Conventional mode-selection methods such as distributed Bragg reflector (DBR) lasers \cite{DBR,siddharth2025ultrafast}, distributed feedback (DFB) lasers \cite{DFB,soda1987stability, Liang:21}, and Vernier-ring avities \cite{Wan:19,wan2020directly,Malik:20} rely on fixed, geometry-defined feedback from etched or regrown gratings. While effective, these approaches often involve complex fabrication steps and lack adaptability, as the resonance condition is permanently set by the cavity geometry. 

Dynamic population gratings (DPGs) provide a fundamentally different mode-control mechanism. A DPG forms when a standing-wave field induces a periodic modulation of the carrier density, creating a self-aligned, optically driven Bragg grating that tracks the lasing field \cite{piccardo2018time,PhysRevA.109.063113}. The strength and stability of this carrier grating depend on the properties of the gain medium, such as carrier diffusion, carrier lifetime, and the density of available electronic states. Strong DPGs have been observed in systems that support pronounced carrier modulation, such as semiconductor structures with high conduction electron densities \cite{vasil2016pulse}  and rare-earth–doped fibers with long-lived excited electronic states  \cite{stepanov2008dynamic, Xu:12}. While DPGs have been widely explored in fiber lasers \cite{poozesh2018single,lobach2017open}, their potential for integrated semiconductor lasers remains largely unexplored because conventional gain materials do not easily sustain the strong carrier gratings required for this effect. 

 Quantum dots (QDs) offer a unique advantage for realizing DPGs on an integrated platform. Their three-dimensional carrier confinement leads to greatly suppressed lateral carrier diffusion compared with quantum wells (QWs) \cite{shang2021perspectives,Shaw_2003,norman2019review,PhysRevLett.98.153903}. This reduced diffusion not only mitigates surface recombination and defect sensitivity \cite{10.1063/1.124003, strand1997reduced,Wei:23,Huang:23}, but also enables a high-contrast, stable carrier grating necessary for strong DPG formation. 
 
\begin{figure*}[t]
\centering
\includegraphics[width=0.9\textwidth]{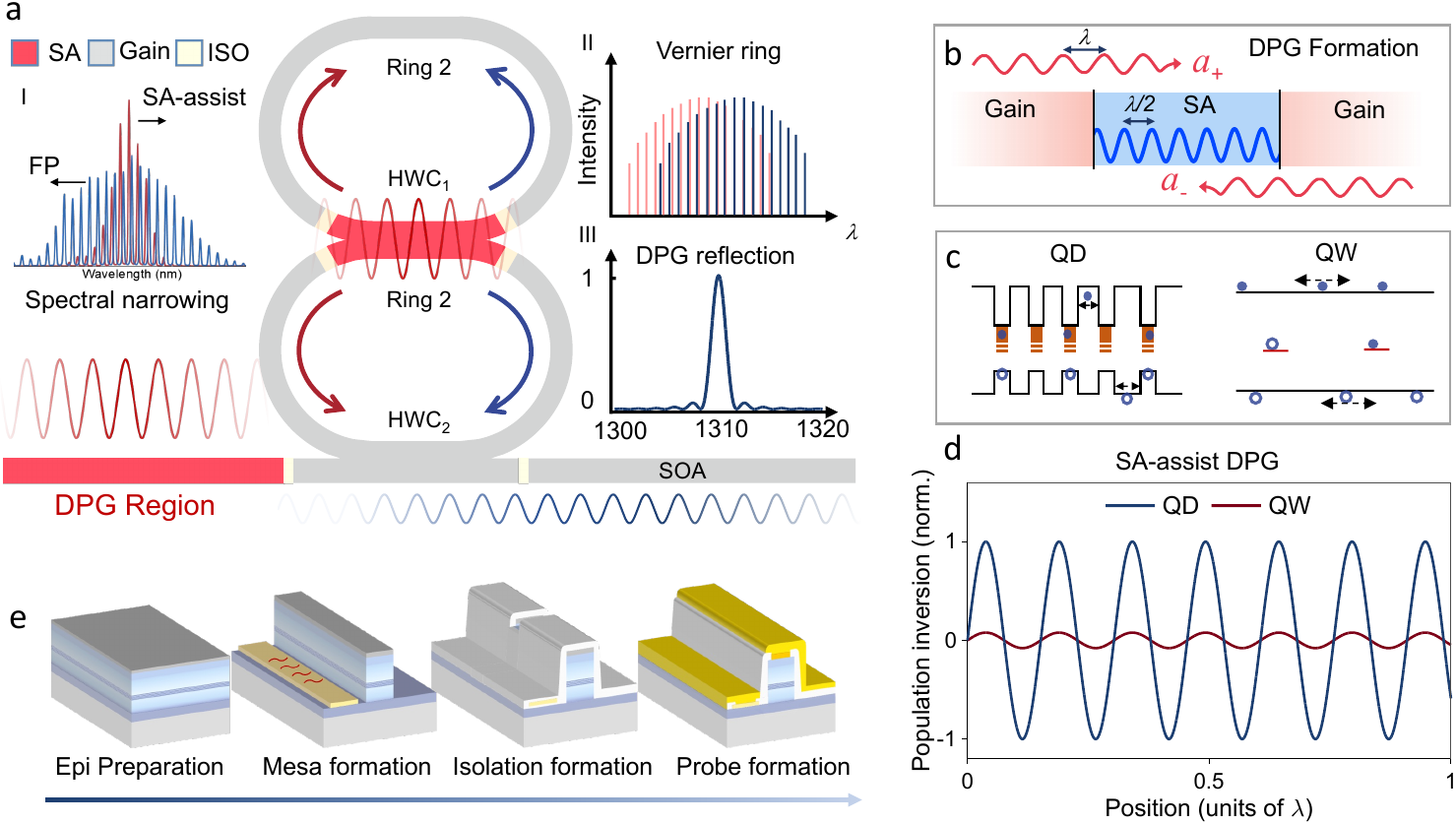}
\caption{
\textbf{Device concept, structure, and principle of operation.} 
(a) Schematic of the dual-ring laser with spatially separated gain section (gray) and reverse-biased SA section (red). The SA introduces nonlinear absorption for mode suppression (I), the detuned rings provide Vernier filtering for coarse selection (II), and the standing-wave–induced DPG supplies reflective feedback that reinforces the dominant mode (III). SA, saturable absorber; ISO, isolation; SOA, semiconductor optical amplifier; DPG, dynamic population grating.
(b)  Illustration of the DPG formation by counter-propagating waves in the SA regions, where the SA enhances the standing-wave induced carrier modulation. 
(c) Lateral carrier diffusion and (d) Calculated carrier-grating profiles in QD and QW active media.
(e) Brief fabrication process flow of QD laser.}
\label{fig1}
\end{figure*}

In this work, we integrate reverse-biased saturable absorbers (SAs) into the QD laser cavity. The SA introduces nonlinear, mode-dependent loss that preferentially suppresses weaker modes while negligibly affecting the dominant mode \cite{PhysRevLett.124.133901,cheng1995stable,PhysRevLett.116.043901}. When a standing-wave pattern forms in this region, the SA enhances the carrier-density modulation, allowing a strong DPG to form and provide  adaptive, wavelength-selective feedback \cite{poozesh2018single,lobach2017open}.  
 Using this mechanism, both single- and dual-ring QD lasers achieve robust SLM operation. The dual-ring Vernier laser (Fig.\,\,\ref{fig1}a), reinforced by DPG-assisted feedback,  achieves a high side-mode suppression ratio (SMSR) of  52.6 dB and a wide wavelength tuning range over 46 nm.  The ultra-low linewidth enhancement factor (LEF) of QDs further provides high feedback tolerance \cite{8719966}, supporting stable lasing under external reflections up to $-10$ dB and isolator-free transmission at 32 Gbps. These results show that combining QD material properties with engineered intracavity loss enables a simple and scalable route to compact, widely tunable, and feedback-resilient semiconductor lasers for large-scale photonic integration.
 
\textit{Dynamic population grating}—To understand the physical origin of the strong mode-selective feedback observed in our QD lasers, we begin by analyzing the mechanism of DPGs formed in the saturable absorber section. DPG is a well-known effect in fiber lasers and has been used successfully to achieve single-mode and narrow-linewidth operation \cite{poddubrovskii2022fiber,lobach2017open}. When the lasing field forms a standing-wave pattern in the SA section of the cavity, a population grating is created.  This grating reflects the carrier density modulation produced by spatial hole burning (SHB) in response to the standing wave \cite{Paschotta:97,Zayhowski:90,PhysRevA.75.031802}, as shown in Fig.\,\ref{fig1}b. For a monochromatic electric field, the spatial period of the grating is half the wavelength of the lasing field. SHB introduces a spatial variation in gain, and this gain modulation, linked to refractive index through the Kramers-Kronig relations, translates into a periodic refractive index modulation \cite{opacak2019theory}. As a result, the carrier grating functions as a Bragg grating that reflects light within a narrow spectral band centered at the wavelength determined by the standing wave. This narrow-band reflection provides effective mode filtering that suppresses lower-intensity side modes and enables stable single-mode lasing.

When the SA is illuminated by two counter-propagating monochromatic waves with wave-vector $k_0$, the carrier density in the SA follows the diffusion equation derived from the Maxwell-Bloch equations:
\begin{equation}
    \frac{\partial N}{\partial t}=D\frac{\partial^2N}{\partial z^2} - \gamma_1(N-N_0) - A \mathrm{sin}^2(k_0z)
,\end{equation}
where $D$ is the carrier diffusion coefficient, $\gamma_1$ is the carrier decoherence rate, $A$ is a constant dependent on the total carrier density and the intensity of the standing wave, and $z$ is the coordinate along the SA section. Solving the carrier-diffusion equation for the steady-state solution yields a carrier density consisting of a constant term $N_c$ and a spatially-varying grating component:
\begin{equation}
    N\propto N_c+N_g \mathrm{cos}(2k_0z),
\end{equation}
where $N_g$ is the amplitude of the carrier grating. Crucially, $N_g$ is inversely proportional to the carrier diffusion coefficient $D$:
\begin{equation}
    N_g\propto\frac{1}{1+4Dk_0^2T_1},
\end{equation}
where $T_1=1/\gamma_1$ is the carrier lifetime. 

This relation shows that the grating strength increases as diffusion decreases; low diffusion allows the standing-wave–induced carrier modulation to persist, while high diffusion washes it out. Because the refractive-index modulation is directly proportional to the grating amplitude ($\delta n \propto N_g$), the resulting Bragg-reflection strength scales as $\delta n^2$. Thus, even moderate changes in carrier diffusion can produce large variations in DPG-assisted feedback. This scaling explains an inherent limitation of conventional QW active regions. QWs typically exhibit lateral diffusion coefficients on the order of $D=60 \ \mathrm{cm^2/s}$ \cite{fiore2004}, which strongly suppresses $N_g$ and leads to weak carrier gratings. In contrast, QDs have much lower diffusion coefficients, often as small as $D=4 \ \mathrm{cm^2/s}$ \cite{kosogov1996structural, 5642656}. This order-of-magnitude reduction in $D$ enables a significantly stronger carrier grating and therefore a much larger DPG-induced reflection that is estimated to be more than 20 dB stronger in a QD medium than in a comparable QW device for a 100 $\mu$m-long SA section. 

These trends are illustrated in Fig.\,\ref{fig1}c and Fig.\,\ref{fig1}d. Fig.\,\ref{fig1}c compares the lateral diffusion behavior of QD and QW materials, emphasizing the strong carrier localization in QDs. Fig.\,\ref{fig1}d shows the calculated  carrier grating profiles, where the QD active region exhibits a high-contrast carrier modulation, while the QW region shows only a shallow grating. These results clearly demonstrate that QDs inherently satisfy the low-diffusion condition required for strong DPG formation, whereas QWs do not.

To quantify this effect, we analyze how this periodic index perturbation couples forward- and backward-propagating fields inside the SA. Letting the probe field be expressed as $a_+$ and $a_-$, the evolution of these components can be described by the coupled-mode equations:

\begin{gather}
    \frac{da_+}{dz}=-(\alpha+ik_p)a_+ + i\frac{\delta n}{n_B}k_0\cos(2k_0z)a_-, \\ 
    \frac{da_-}{dz}=(\alpha+ik_p)a_- - i\frac{\delta n}{n_B}k_0 \cos(2k_0z)a_+,
\end{gather}
where $\alpha$ is the propagation loss in the SA section ($\approx 1.2 \ \mathrm{dB/mm}$), $k_p$ is the wave-vector of the probe field and $n_B$ is the background refractive index. Solving these equations yields the reflection coefficient as a function of detuning $\Delta k=k_p-k_0$ \cite{drobyshev2019spectral}:
\begin{equation}
    R(\Delta k) \propto \delta n^2 \left| \frac{1-e^{-2(\alpha+i\Delta k)L}}{\alpha+i\Delta k}  \right|^2,
\end{equation}
where $L$ is the SA length. This expression shows several important features. First, the grating bandwidth is mainly set by $L$, which means, a longer SA produces a narrower reflection peak. Second, the reflection amplitude is limited by the SA propagation loss $\alpha$, so operating the SA under reverse bias (which lowers the effective loss for the dominant mode) is beneficial for maximizing DPG-assisted feedback.

Most importantly, because the refractive-index modulation scales as $\delta n \propto N_g$, and $N_g$ itself depends inversely on the diffusion coefficient $D$, the reflection strength follows 

\begin{equation}
    R\propto\ N_g^2 \propto\frac{1}{(1+4Dk_0^2T_1)^2},
\end{equation}

This scaling confirms that materials with low carrier diffusion naturally provide significantly stronger DPG-assisted feedback, consistent with the QW–QD comparison shown earlier in Fig.\,\ref{fig1}c and Fig.\,\ref{fig1}d. The calculated reflection spectrum at 1310 nm for a QD-gain DPG formed within a 100-$\mu$m-long SA section is shown in Fig. 1a(III).

\begin{figure*}[t]
\centering
\includegraphics[width=0.9\textwidth]{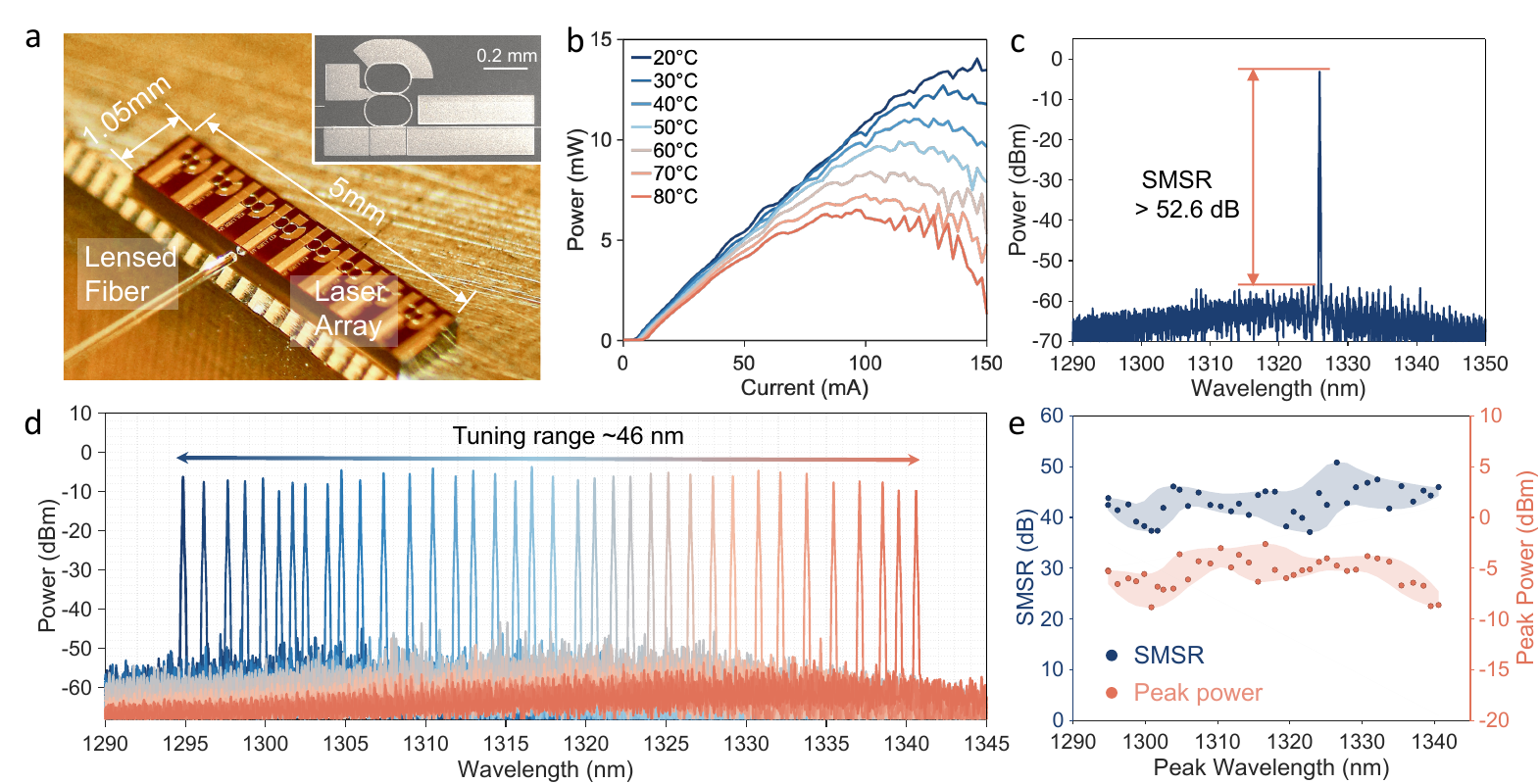}
\caption{\textbf{Device performance of the dual-ring QD laser with DPG.} 
(a)  Optical micrograph of the fabricated tunable laser array, consisting of nine devices on a 5 mm × 1.05 mm bar. Inset: SEM image of a single dual-ring laser.
(b) LIV curves measured at various chip temperatures, confirming continuous-wave lasing up to $80\,^{\circ}\mathrm{C}$. 
(c) Representative single-mode spectrum with SMSR exceeding 52.6~dB. 
(d) Measured tuning characteristics, demonstrating a wavelength tuning range of \~46 nm.
(e) Extracted SMSR and peak power as a function of lasing wavelength.
}
\label{fig2}
\end{figure*}
\textit{Tunable laser with DPG}—Building on the above insights, we now apply the DPG mechanism to both single-ring and dual-ring configurations. 

In the single-ring structure, robust single-mode lasing with an SMSR over 46 dB is achieved (Fig. S4a), with side modes effectively suppressed by the DPG formed in the cavity. This serves as a proof-of-concept demonstration that DPG-enabled mode selection can operate reliably in an integrated QD laser platform. 

To achieve wide tunability while maintaining strong mode selectivity, we incorporate the DPG mechanism into a dual-ring structure. The device consists of a bus waveguide coupled to two cascaded microrings with slightly different circumferences, as shown in Fig.\,\ref{fig1}a. The two rings are connected through a half-wave coupler (HWC), which introduces a $\pi $ phase difference between the self-coupling and cross-coupling coefficients to enable synchronous power transfer. A second identical HWC couples the cascaded rings to the bus waveguide, ensuring matched phase conditions and symmetric coupling at both interfaces \cite{He:08}.    

In a Vernier laser, the overall tuning span is determined by the enlarged FSR of the combined cavities, which increases as the cavity-length difference decreases.  However, the cavity-length mismatch cannot be reduced indefinitely. As the two FSRs approach one another, the modal gain contrast between adjacent Vernier resonances diminishes, reducing the threshold difference between the main and side modes and thereby lowering the SMSR. In our device, the DPG provides an additional mechanism of mode selectivity that compensates for this limitation. With the DPG reinforcing the dominant longitudinal mode, the Vernier effect can serve primarily as a coarse spectral filter. This allows us to operate with a smaller detuning of 1.7\%, realizing an enlarged FSR of 45 nm. The corresponding transmission spectrum is shown in Fig.\,S3(a).   

 Fig.\,\ref{fig1}e outlines the fabrication process flow of the QD lasers, including epitaxial wafer preparation,  mesa definition, electrical isolation, and probe metal formation. The optical micrograph  and the scanning electron microscopy (SEM) image of fabricated tunable laser arrays are shown in Fig.\,\ref{fig2}a. The lasers are then mounted onto a temperature-controlled copper plate with a thermoelectric cooler (TEC) for measurement. The injection current of the amplifier part ( $I_{SOA}$ )  is swept, while both injection currents on  $R_1$,  $R_2$ along with $HWC_2$ are fixed at 10 mA using two independent current sources to make these sections nearly transparent.  The first coupler $HWC_1$ is biased at $-2.5$ V to operate as an SA, and the left end of the Fabry–Pérot cavity remains unbiased, naturally functioning as an SA. The light–current (L–I) characteristics measured at various heat-sink temperatures are shown in Fig.\,\ref{fig2}b. The laser delivers a continuous-wave (CW) output power exceeding 14.5 mW, while maintaining stable operation up to $80\,^{\circ}\mathrm{C}$, highlighting the excellent thermal robustness of the QD gain medium. 
 
 A  representative SLM lasing spectrum at $1320.4~\mathrm{nm}$ is shown in Fig.\,\ref{fig2}c, exhibiting an SMSR exceeding $52.6~\mathrm{dB}$. This highly purified spectrum highlights the effectiveness of the SA-assisted DPG in suppressing longitudinal modes competition and enforcing SLM operation. The wide tunability enabled by the Vernier configuration is shown in Fig.\,\ref{fig2}d. By varying the injection currents, the lasing wavelength can be tuned from $1295$ to $1341~\mathrm{nm}$, corresponding to a tuning span of $46~\mathrm{nm}$ while maintaining a SMSR higher than $38~\mathrm{dB}$. Because the QD gain spectrum redshifts at a rate of 0.35 nm/$^{\circ}\mathrm{C}$ with increasing  temperature \cite {Wan:19}, the tuning range can be further extended by changing the TEC temperature.  The extracted SMSR and the peak powers of selective wavelengths are shown in Fig.~\,\ref{fig2}e. The results indicate that the SA-assisted DPG could be adaptively changed by the center wavelength and consistently enable high SMSR and stable output power throughout the entire tuning range. \\
\begin{figure*}[t]
\centering
\includegraphics[width=0.9\textwidth]{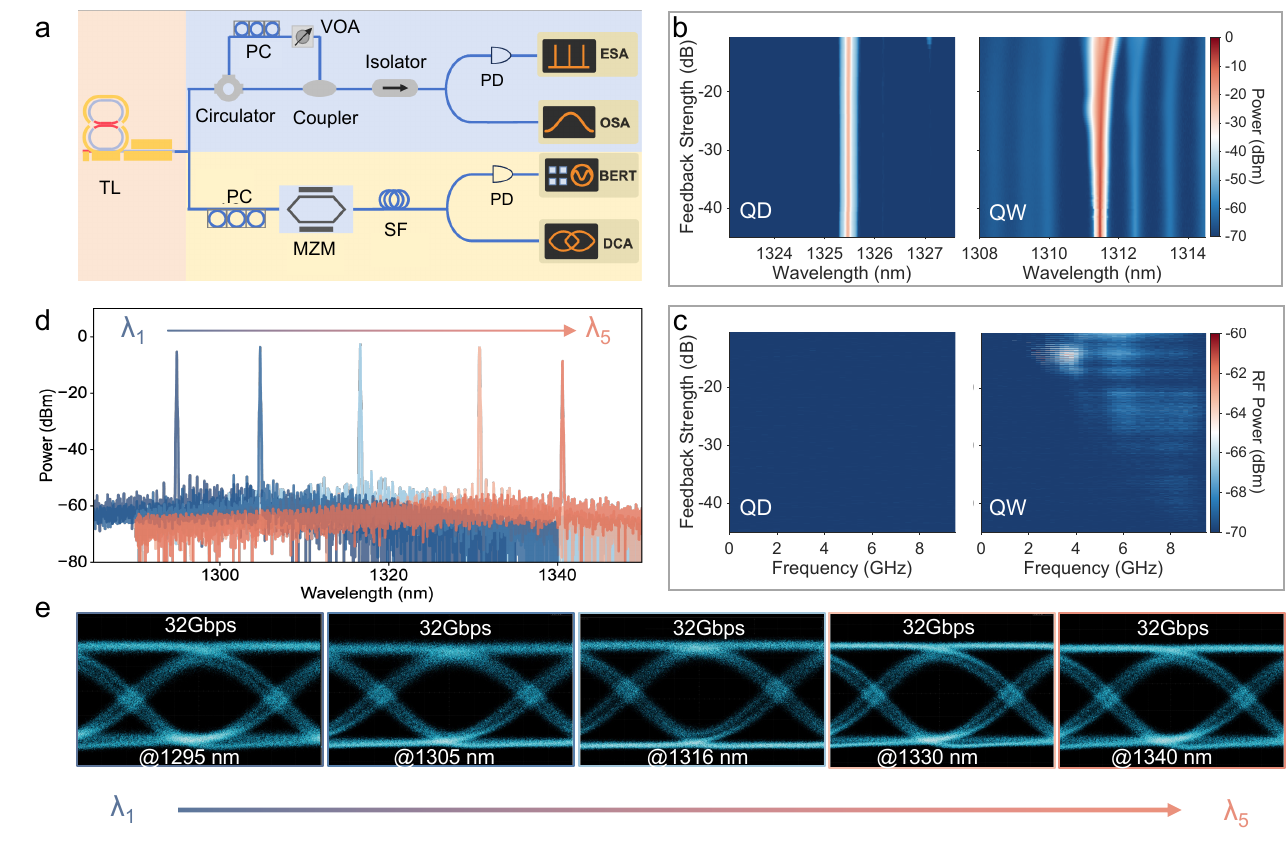}
\caption{\textbf{Feedback dynamics and isolator-free transmission of the QD laser.} 
(a) Experimental setup of feedback dynamics and isolator-free transmission link.
(b) Spectrogram of the QD and QW lasers under increasing external optical feedback, showing robust single-mode operation of the QD laser up to $-10.6$~dB feedback. 
(c) RF spectra of the QD and QW lasers under the same conditions, confirming the absence of coherence collapse of the QD laser. 
(d) Isolator-free emission spectra of the QD tunable laser at five representative wavelengths. 
(e) Eye diagrams recorded at 32~Gbps for five representative wavelengths of the QD laser without an isolator, showing clear and open eyes with error-free operation.}
\label{fig3}
\end{figure*}
\textit{Feedback insensitivity and isolator-free communication system}—The stable SLM operation and consistent output power across the entire tuning range demonstrate that the SA-assisted DPG provides robust mode control. To further assess the reliability of the tunable QD laser under practical operating conditions, we evaluate its sensitivity to external optical feedback (EOF)and its performance in a fully isolator-free communication system.

Fig.\,\ref{fig3}a depicts the experiment configuration for EOF and isolator-free data transmission. The QD and QW lasers are temperature stabilized on a TEC stage. The output is coupled into a lensed fiber and routed through a three-port fiber-optic circulator, which directed the signal into a 90/10 fiber coupler. The $10\%$ output arm is used for optical spectral analysis and relative intensity noise (RIN), switched via a fiber-optic switch. The $90\%$ output arm passes through a variable optical attenuator (VOA) to change the feedback strength $ r_{ext}$, defined as the ratio between the optical power returned into the laser cavity and the emitted free-space power. The attenuated signal then passes through a manual polarization controller (PC) to compensate for polarization rotation in the fiber and maximize the observable effects of EOF. The signal is subsequently coupled back into the laser cavity via the circulator to form the controlled EOF loop. The insertion losses of the fiber lens (2.9 dB), circulator (0.57 dB), couplers  (0.5 dB), and connectors were carefully calibrated to ensure accurate determination of the feedback strength.

The feedback dynamics were evaluated in both the optical and RF domains. Fig.\,\ref{fig3}b presents the spectrograms of  QD and QW lasers as the feedback strength increased from $-45~\mathrm{dB}$ to $-10.6~\mathrm{dB}$.  In our measurements,  $-10.6~\mathrm{dB}$ represents the strongest feedback achievable after accounting for all link losses. Remarkably, the QD laser preserves a single narrow emission line throughout the entire feedback range. In contrast, the QW laser undergoes clear signs of feedback-induced degradation: spectral broadening, side-mode activation, and coherence collapse when the feedback exceeds $-40~\mathrm{dB}$. Beyond this critical feedback level ($ r_{ext}$), the QW laser shows a slight redshift and broadening of the longitudinal modes, consistent with the onset of coherence-collapse dynamics. 

This behavior is evident in the RF spectrum. The QD laser shows no evidence of nonlinear oscillations or excess RF noise as the feedback increases. However, the QW lasers present undamped relaxation oscillations that evolve into chaotic behaviors as the feedback increases. This stark difference originates from the intrinsic carrier dynamics of the two gain media \cite{dong2024turnkey,shi2025exploring,PhysRevA.103.033509, Cui:24}. In QW lasers, the LEF $\alpha $ couples carrier-induced refractive index fluctuations ($\Delta n$) to gain variations ($\Delta g$), making the cavity phase highly susceptible to perturbations from external reflections. In QD lasers, the discrete density of states and symmetric gain spectrum yield a much smaller $\alpha$ factor. Together with strong gain compression and ultrafast intraband relaxation, these features suppress carrier-induced index fluctuations and weaken the coupling between amplitude and phase noise \cite{Su_2005,806834,1263967,Rasmussen2019PRL}. Consequently, QD lasers exhibit superior tolerance to EOF. 

Leveraging this intrinsic robustness, we implement an isolator-free transmission system directly driven by the QD laser (Fig.3a). The setup employs a commercial external lithium–niobate Mach–Zehnder modulator (MZM), and we evaluate transmission at five representative lasing wavelengths of the tunable QD device, as presented in Fig.\,\ref{fig3}d. The QD laser output is first coupled into a PC and then directly fed into the MZM. The modulated signal is transmitted through a standard single-mode fiber (SF) and analyzed by a bit-error-rate tester (BERT) and a digital communication analyzer (DCA) for high-speed eye-diagram measurements. This configuration enables performance benchmarking across multiple lasing wavelengths without using an optical isolator.

The corresponding eye diagrams, recorded at 32~Gbps (Fig.\,\ref{fig3}e), show clear and open eyes across all five tested wavelengths without any observable penalty from the absence of an optical isolator. For comparison, we also measure the system-level performance using a commercial bulk laser with an integrated isolator. The resulting eye diagram (Fig.\,S4e) is nearly identical to that of our isolator-free QD laser.  These results demonstrate that QD lasers combine superior feedback resilience with excellent high-speed modulation performance, offering a compact, energy-efficient, and isolator-free light source solution for silicon photonic transceivers and reconfigurable optical networks.  \\
\begin{figure}[t]
\centering
\includegraphics[width=\columnwidth]{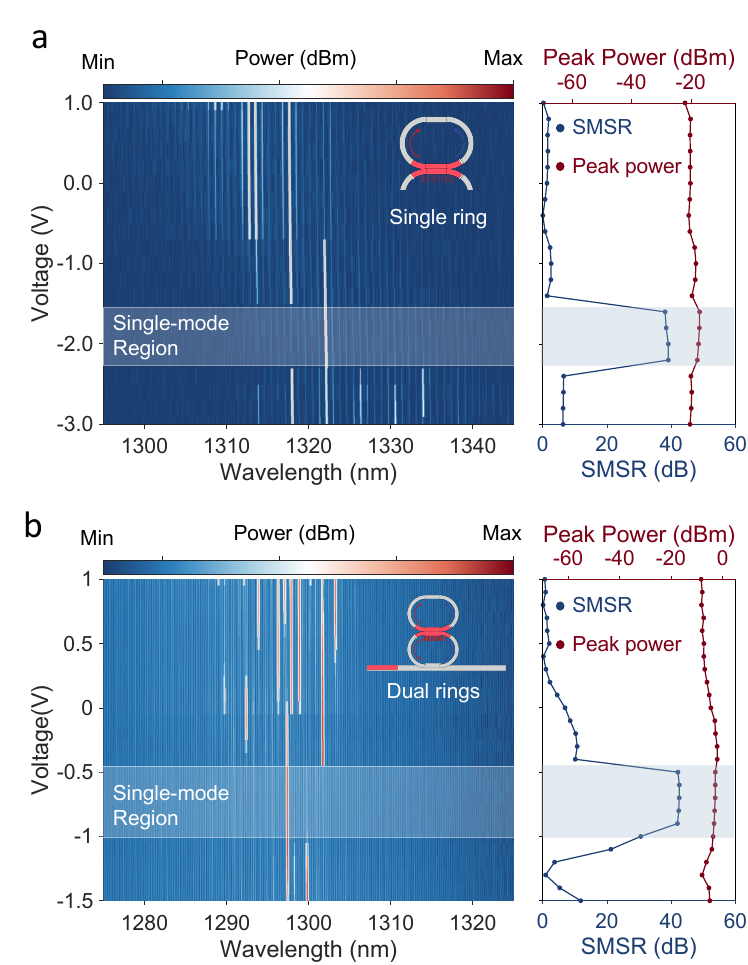}
\caption{\textbf{Single-ring and dual-ring lasers with HWC and SA-assisted DPG operation.} 
(a) Voltage-controlled spectral evolution of the single-ring laser at 40 mA, showing the emergence of a stable single-mode window confirmed by the extracted SMSR and peak-power curves.
(b) Voltage-dependent lasing spectra of the dual-ring Vernier laser with the currents of the two rings and the SOA are fixed at $22~\mathrm{mA}$, $26~\mathrm{mA}$, and $60~\mathrm{mA}$, respectively.
\label{fig4}}
\end{figure}
\textit{DPG generation and discussion}— To better understand the physical origin of this stability, we examine the formation of DPGs in our devices and analyze their role in enforcing single-mode operation. Here, we use different laser structures and pumping conditions to experimentally map the presence of DPGs and the range over which they provide effective mode filtering. 

First, we investigate a single-ring laser with a HWC to study voltage-dependent DPG behavior. Fig.\,\ref{fig4}a shows the evolution of the lasing spectrum as the SA bias is swept at a fixed gain current of 40 mA. A stable single-mode region is observed throughout a wide voltage range. The extracted SMSR and peak output power as a function of SA bias are presented on the right side of Fig.\,\ref{fig4}a.  A stable DPG window between $-2.25$ and $-1.50~\mathrm{V}$ is clearly identified, within which the laser maintains high-SMSR and stable single-mode lasing. As the bias is increased toward +1 V,, the SA becomes increasingly transparent that reduces its ability to suppress side modes.  Competing modes reappear as the DPG shifts away from the lasing wavelength. When the bias is further reversed beyond $-2.5$ V, the SA introduces excessive loss to all modes, including the dominant one, reducing net gain and the effectiveness of the DPG-assisted feedback. 

We also perform current scanning under a fixed voltage bias of –2.0 V (Details provided in Supplementary materials). As the gain current is swept from 0 to 80 mA, the spectrogram shows a clear single-mode lasing channel centered near the lasing wavelength, indicating that the DPG provides narrowband distributed feedback over a wide dynamic range. For comparison, we perform an identical spectrogram mapping with the SA forward-biased at 10 mA, and we observe that the HWC section becomes nearly transparent. As shown in Fig. S4(b), multimode lasing dominates across the entire current range, confirming that the absorptive SA region is essential for generating the DPG. 

We further investigated the DPG effect for the dual-ring Vernier laser. Fig.\,\ref{fig4}b shows the spectrogram as the SA bias is swept from $-1.5$ to $+1.0~\mathrm{V}$, with the currents of the two rings and the SOA fixed at $22~\mathrm{mA}$, $26~\mathrm{mA}$, and $60~\mathrm{mA}$, respectively. Similar to the single-ring lasers, a bias-defined DPG window emerges from $-1.02$ V to $-0.45$ V and continuously aligns with the lasing mode. Within this window, the SMSR remains above $35~\mathrm{dB}$,  while the output power stays nearly constant at $-2$ to $-4~\mathrm{dBm}$. Outside this voltage range, reduced absorption or excessive loss disrupts DPG formation, leading to the reappearance of multimode operation. These observations in both single-ring lasers and dual-ring lasers demonstrate that the SA provides an electrically tunable DPG feedback, which enables dynamic, wavelength-selective feedback. This mechanism establishes a controllable single-mode window and offers precise longitudinal-mode selection across a wide range of operating conditions.

\textit{Conclusion}—In summary, we have theoretically analyzed and experimentally validated a DPG approach for achieving robust single-mode operation in QD lasers. The discrete energy states and low lateral carrier diffusion inherent to QDs enable a strong standing-wave-induced carrier grating to form in a reverse-biased SA. This provides self-aligned, wavelength-selective feedback without requiring regrowth, etched gratings, or other complex fabrication steps. By incorporating this mechanism into single- and dual-ring lasers, we demonstrate stable single-longitudinal-mode lasing with SMSR up to 52.6 dB and a continuous tuning range of 46 nm. The devices maintain CW operation up to $80\,^{\circ}\mathrm{C}$ and preserve single-mode lasing under EOF as strong as $–10.6$ dB, enabling isolator-free data transmission at 32 Gbps with clear, open eye diagrams. These results establish DPG-enabled QD lasers as a simple route to tunable, feedback-tolerant on-chip light sources. More broadly, the ability to electrically control a self-formed carrier grating introduces a new degree of freedom for mode control in semiconductor lasers and provides a scalable pathway for integrated photonic systems in communications, sensing, and reconfigurable optical networks.

\textit{Acknowledgments}—The authors acknowledge the support of the KAUST Core Labs for device fabrication.

\textit{Funding}—This publication is based upon work supported by the King Abdullah University of Science and Technology (KAUST) under Award No. RFS-TRG2024-6196, ORFS-CRG12-2024-6487, RFS-OFP2023-5558, and FCC/1/5939.

\textit{Data availability}—All data are available from the corresponding author upon reasonable request. \\

\bibliography{sample}

\end{document}